\documentclass[aps,prb,twocolumn,superscriptaddress,floatfix,longbibliography]{revtex4-2}
\usepackage{graphicx}
\usepackage{bm}
\usepackage{wrapfig}
\usepackage{color}
\usepackage{natbib}
\usepackage{hyperref}
\usepackage{amssymb}
\usepackage{amsmath}
\usepackage{geometry}
\usepackage{comment}

\hypersetup{
colorlinks=true,
urlcolor= blue,
citecolor=blue,
linkcolor= blue,
}

\newif\ifptitle
\newif\ifpnumber
\newcounter{para}

\ptitletrue  
\pnumbertrue  

\def\XXint#1#2#3{{\setbox0=\hbox{$#1{#2#3}{\int}$}
     \vcenter{\hbox{$#2#3$}}\kern-.5\wd0}}

\begin{document}

\title{Hall effect induced by topologically trivial target skyrmions}

\author{Tan Dao}
\affiliation{Department of Physics, Harvard University, Cambridge, MA 02138, USA }
\affiliation{Department of Physics and Astronomy, University of New Hampshire, Durham, New Hampshire 03824, USA }

\author{Sergey S. Pershoguba}
\affiliation{Department of Physics and Astronomy, University of New Hampshire, Durham, New Hampshire 03824, USA }

\author{Jiadong Zang}
\email[ ]{Jiadong.Zang@unh.edu}
\affiliation{Department of Physics and Astronomy, University of New Hampshire, Durham, New Hampshire 03824, USA }
\affiliation{ Materials Science Program, University of New Hampshire, Durham, New Hampshire 03824, USA}

\date{\today}

\begin{abstract}
Electrons moving through a noncoplanar magnetic texture acquire a Berry phase, which can be described as an effective magnetic field. This effect is known as the topological Hall effect and has been observed in topological spin textures. Motivated by recent experimental realizations, here we study the Hall effect in a nontopological magnetic texture known as a target skyrmion. We start from a simplified semiclassical picture and show that the Hall signal is a nonmonotonic function of both the electronic energy and target skyrmion radius. That observation carries over to the fully quantum mechanical treatment in a Landauer-B\text{\"u}ttiker formalism in a mesoscopic setting. Our conclusion challenges the popular opinion in the community that the Hall effect in such structures necessarily requires a nonzero skyrmion number.  
\end{abstract}

\maketitle

\section{Introduction}
In contrast to the classical Hall effect \cite{hall1879new} induced by a nonzero external magnetic field, 
the anomalous Hall effect (AHE) represents a component of the Hall signal proportional to the magnetization of a magnet \cite{karplus1954hall, nagaosa2010anomalous}. Therefore it can exist even in the absence of the external magnetic field. In frustrated chiral magnets, there is an additional contribution to AHE arising from the noncoplanar structure of spin magnetic moments \cite{taguchi2001spin, fujishiro2021giant, ghimire2018large, yang2020giant, ohuchi2018electric}. In these materials, electrons acquire a Berry phase, therefore generating an anomalous velocity which can be understood as the Lorentz force of an effective magnetic field in momentum space \cite{karplus1954hall,nagaosa2010anomalous,taguchi2001spin, fujishiro2021giant}. More interestingly, in the adiabatic regime where itinerant electrons are strongly coupled to the spins, the Berry curvature in real space can give rise to the AHE effect \cite{xiao2010berry,bruno2004topological, do2009skyrmions}. This effect is also known as the topological Hall effect (THE). As the real space Berry phase is directly related to the topology of the spin texture, the THE is often used as evidence of nontrivial spin texture \cite{hamamoto2015quantized, kurumaji2019skyrmion, neubauer2009, gerber2018interpretation, vistoli2019giant, kanazawa2011large}. One notable example of nontrivial spin texture is the magnetic skyrmion, whose topological charge (Q = 1) is defined as the integer winding number of spin $\bm{S(r)}$ on a unit sphere \cite{nagaosa2013topological}. 

Since the discovery of skyrmions in MnSi\cite{neubauer2009}, the role of topology and magnetization has increasingly expanded for its potential use in dense and robust data storage, as well as racetrack memory for quick data access \cite{finocchio2016magnetic, luo2021skyrmion, fert2017magnetic}. However, chiral spin textures that are classified as topologically trivial have remained underexplored in theory and experiment. In particular, the target skyrmion whose texure is a concentric ring of one skyrmion inside another skyrmion with reversed spins (see Fig.\ref{fig:target_skyrmion}a) can be moved linearly with a current, making it promising for racetrack memory applications \cite{tang2021magnetic, zhang2016control, kolesnikov2018skyrmionium, zhang2018real, shen2019current}. Currently, target skyrmion are being identified by techniques such as magnetic force microscopy, Lorentz transmission electron miscroscopy, X-ray photoelectron spectroscopy, and electron hollography, which can not be integrated to electronic devices \cite{tang2021magnetic, zhang2018real, seng2021direct}. The topological Hall effect, however, can be used to quickly detect presence of spin texture by measuring the transverse voltage of devices. Here we show that the topological trivial target skyrmion can give rise to the Hall effect.   

We start with a classical model of electrons traveling in the effective magnetic field of the target skyrmion and show a non-zero transverse current. We then performed quantum transport calculations and observed a sign change in the non-monotonic Hall angle dependence of target skyrmion radii. This indicates the inner and outer skyrmion shells contribute to the Hall angle independently. This result is further supported by the calculation of the Hall angle for each shell; showing that the sum of the Hall angle of each shell is the same as the target skyrmion's Hall angle. We also shows that our results respect the adiabatic approximation. This implies that the Hall effect came from real space topology, which can be interpreted as the THE.   

\section{Target skyrmion configuration} \label{sec:target_sk_conf}
To set the stage, we discuss details of a target-skyrmion texture in this section. We consider a two-dimensional ferromagnet described by a magnetization vector $\bm S(\bm r)$ normalized to unity $|\bm S(\bm r)| = 1$. We may use the following parametrization of the target skyrmion,
\begin{align}
\bm S(\bm r) &= \left(\begin{array}{c}
         \sin\phi(\bm r)\,\sin\theta(\bm r) \\
         \cos\phi(\bm r)\,\sin\theta(\bm r) \\
         \cos\theta(\bm r)
\end{array}\right),  \nonumber
\end{align}
where the polar and azimuthal angles 
\begin{equation}
\begin{aligned}
    \phi(\bm r) &= \tan^{-1}\left(\frac yx\right) - \pi/2 \\
    \theta(\bm r) &= 2\pi - 8\tan^{-1}\left[\exp\left(\frac{4r}r_0\right)\right]
\end{aligned} \label{angles_def}
\end{equation}
are position-$\bm r$ dependent.
 The specific choice of the angle~(\ref{angles_def}) is motivated by the necessity to eliminate spurious electronic scattering at the target skyrmion boundary \cite{ndiaye2017, tang2021magnetic}.



\begin{figure}
    \centering
     \includegraphics[width=1.0\linewidth]{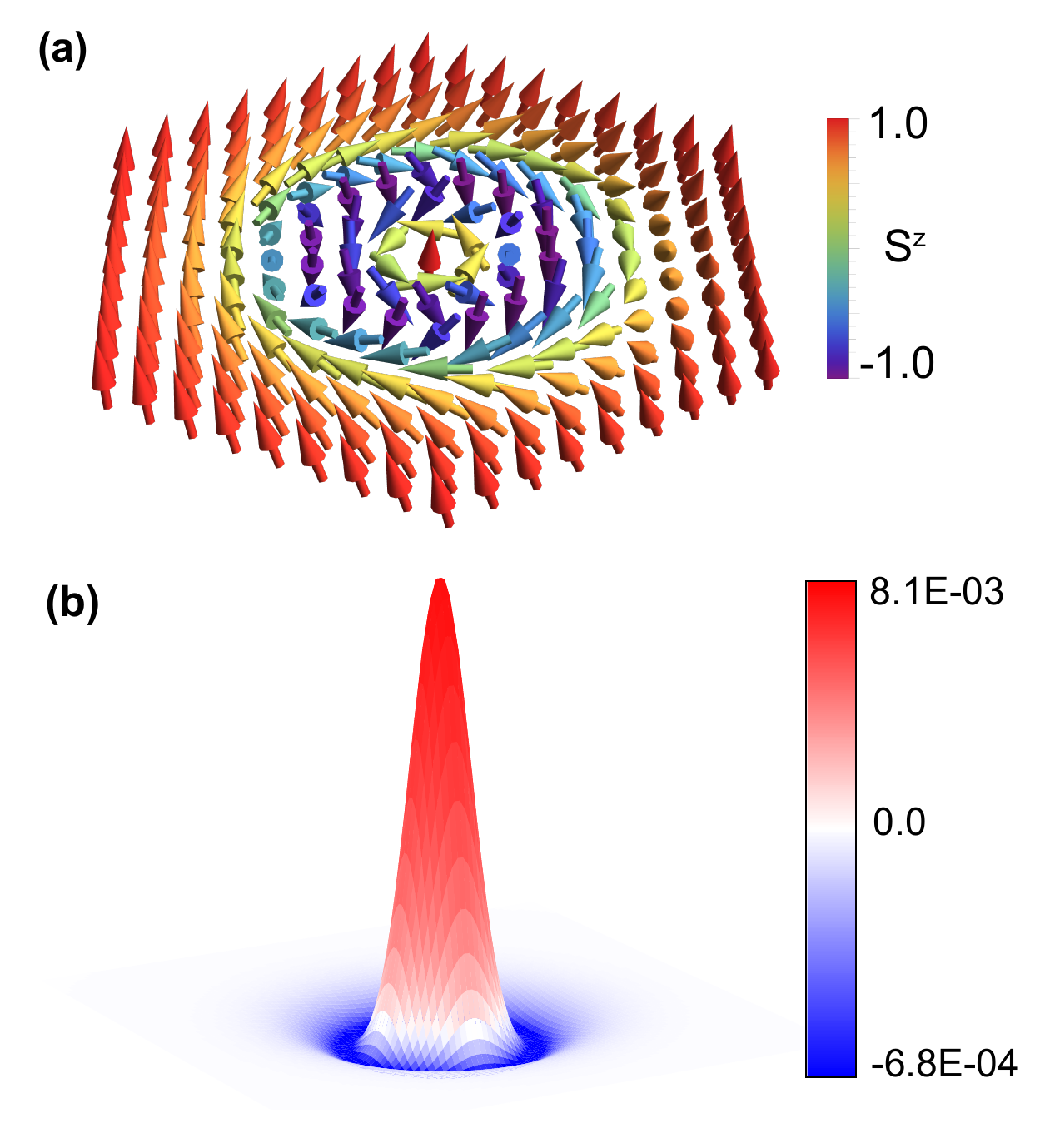}
     \caption{a) Target skyrmion spins configuration. The arrows indicate the spins rotation, and the colorbar indicates $S^z$ magnitude. b) The effective magnetic field of a target skyrmion.} 
     \label{fig:target_skyrmion}
\end{figure}

Assuming that we are working in the adiabatic regime, we may apply the real space Berry phase picture by evaluating the effective magnetic field 
\begin{equation}
B_z(\bm r) = \frac{1}{2}\bm S(\bm r)\cdot[\partial_x \bm S(\bm r) \times \partial_y \bm S(\bm r)]
\end{equation}
The effective magnetic field distribution of the target skyrmion is shown in Fig. \ref{fig:target_skyrmion}b. The inner shell's field is positive and greatest at the center of the target skyrmion due to a larger gradient of the spins at around the center spin. The outer shell's field is negative and is much weaker. The total magnetic field added up to zero which is consistent with $Q = 0$ of the target skyrmion \cite{tang2021magnetic, zhang2016control}. We use the effective field of the target skyrmion for our classical numerical calculations. 


\section{Classical model}

\begin{figure}[hbt!]
    \centering
     \includegraphics[width=1.0\linewidth]{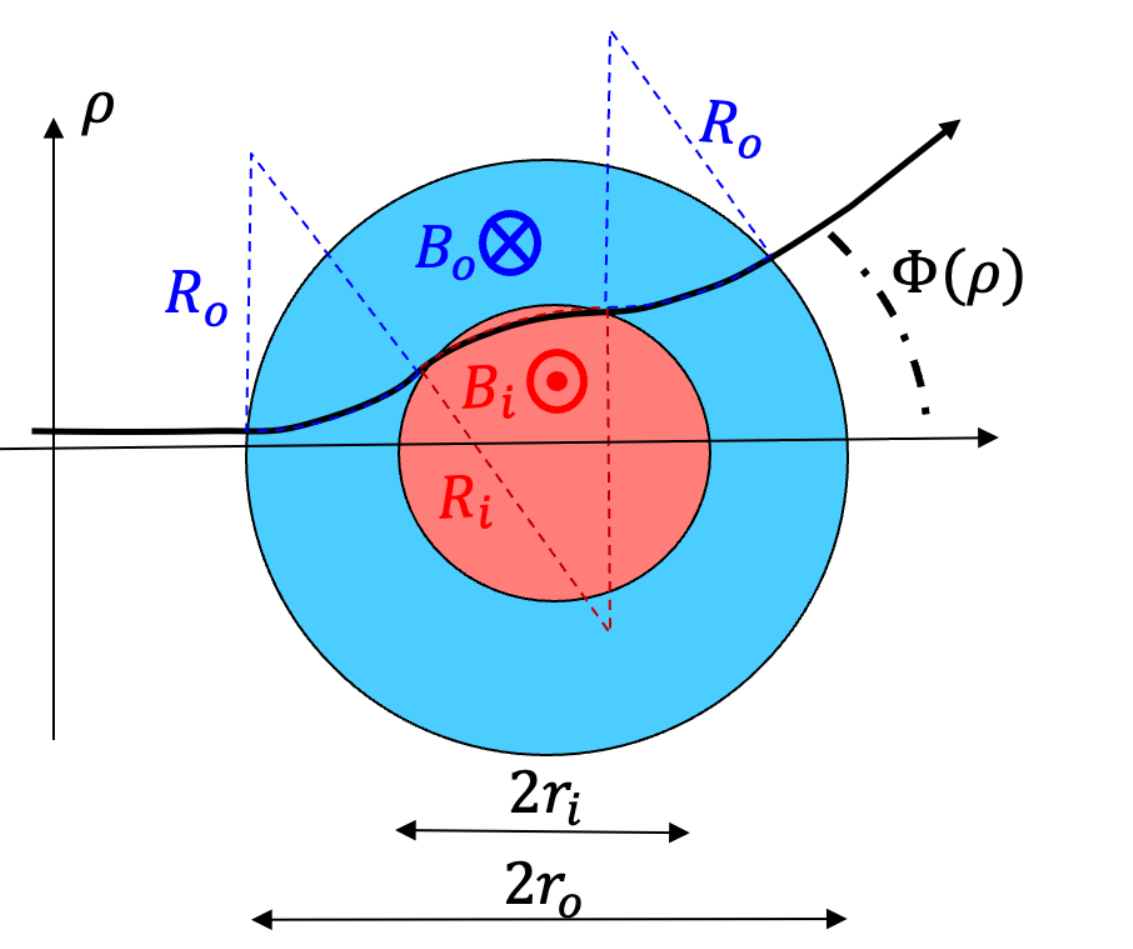}
     \caption{(a) Approximate representation of a target skyrmion. The magnetic fields in the inner (i) circular and outer (o) annular regions have opposite signs [Fig.~\ref{fig:target_skyrmion}(b)]. The opposite Lorentz force bends the corresponding cyclotron trajectories in the corresponding regions in the opposite direction. The scattering angle $\Phi(\rho)$ is given by Eq.~(\ref{scat_angle}).} 
     \label{fig:annulus}
\end{figure}
As a warm-up, we analyze classical scattering of electrons of the target skyrmion. We approximate the target-skyrmion as a two-shell structure shown in the Fig.~\ref{fig:annulus}. The inner (i) circular shell has a radius $r_i$, and the outer (o) annular region has radius $r_o$. The effective magnetic fields have opposite signs $\bm B_i = +\hat{\bm z}\,B_i$ and $\bm B_o = -\hat{\bm z}\,B_o$, which we assume constant within the respective regions. The specific values are constrained by the condition that the total magnetic flux piercing through the structure vanishes, i.e. 
\begin{align}
    B_i \pi r_i^2 = B_o \pi \left(r_o^2-r_i^2\right). \label{equal_fluxes_constraint}
\end{align}
In the presence of the effective magnetic fields $B_i$ and $B_o$, the particles moves along the cyclotron trajectories with radii 
\begin{align}
    R_{i} = \frac{p}{B_{i}}\,\,{\rm and}\,\, R_{o} = \frac{p}{B_{o}}
\end{align}
determined by the magnetic fields in the respective regions $B_i$ and $B_o$ as well as the momentum $p =  \sqrt{2 m E}$. A representative electronic trajectory is shown in Fig.~\ref{fig:annulus}. By matching the circular trajectories in the two regions, we find the scattering angle dependence 
\begin{align}
    &\Phi(\rho) = 2\,{\rm Re}\left\{\,{\rm \,acos}\left[\frac{R_o+\rho}{\sqrt{R_o^2+r_o^2+2\rho R_o}}\right] \right. \nonumber\\ &\qquad- {\rm\,acos}\left[\frac{2R_o^2+2\rho R_o+r_o^2-r_i^2}{2R_o\sqrt{R_o^2+r_o^2+2\rho R_o}}\right]\label{scat_angle}  \\
    &-\left.\,{\rm \,acos}\left[\frac{2R_iR_o-2\rho R_o+r_i^2-r_o^2}{2R_o\sqrt{R_i^2+r_i^2-2\rho R_i+(r_i^2-r_o^2)R_i/R_o}}\right]\right\}.  \nonumber
\end{align}
on the impact parameter $\rho$. 

To get a feeling of Eq.~(\ref{scat_angle}), let us give the low-energy (high-field) limiting cases of  $R_o/r_o \ll 1$ 
\begin{align}
    \Phi(\rho) \approx 2\,{\rm Re}\left\{\,{\rm acos}\left[\frac{\rho}{r_o} + \frac{R_o}{r_o} \left(1 - \frac{\rho^2}{r_o^2}\right) \right]\right\}, \label{low_energy}
\end{align}
where only the first term in Eq.~(\ref{scat_angle}) corresponding to the outer skyrmion contributes, as well as the high-energy (low-field) limit $r_o/R_o \ll 1$
\begin{align}
    \Phi(\rho) \approx 2{\rm Re}\left\{ \frac 1R_o \sqrt{r_0^2-\rho^2} - \frac{R_o+R_i}{R_oR_i} \sqrt{r_i^2-r_0^2}\right\}. \label{high_energy}
\end{align}
Here, the former and the latter terms represent the effect of the outer and inner skyrmions, respectively. The opposite sign of those terms correspond to the opposite direction of the Lorentz force acting in the respective regions. Then, it is instructive to evaluate the skew-scattering Hall (H) component of the differential cross-section (having units of length in 2D)
\begin{align}
    \sigma_{H} = \int_{-\infty}^\infty d\rho \,\sin [\Phi(\rho)]. \label{sk_cross_sect}
\end{align}
In the rest of the paper, we discuss the Hall resistance $R_H$ induced by a target skyrmion, so there is no confusion between the Hall conductivity and Eq.~(\ref{sk_cross_sect}). Evaluating Eq.~(\ref{sk_cross_sect}) for the low-energy (\ref{low_energy}) and the high-energy (\ref{high_energy}) limits, we obtain
\begin{align} 
 \sigma_H &\approx \frac{\pi R_o}2 \,\,\,{\rm for}\,\,\, \frac{R_o}{r_o} \ll 1, \label{cross_sect_low}\\
 \sigma_H &\approx \pi \left[\frac{r_o^2}{R_o}-\frac{(R_i+R_o)r_i^2}{R_oR_i}\right],\,\,\,{\rm for} \,\,\,\frac{r_o}{R_o} \ll 1. \label{cross_sect_high}
\end{align}
It is expected that at low energies, the carriers are deflected by the Lorentz force and are not able to penetrate into the inner shell of the target skyrmion. So the skew-scattering cross-section~(\ref{cross_sect_low}) is determined only by the effective magnetic field of the outer skyrmion. In the high-energy limit, it is interesting to note that $\sigma_H \to 0$ vanishes once the constraint of equal magnetic fluxes~(\ref{equal_fluxes_constraint}) is taken into account. It corresponds to the known result in the literature~\cite{denisov2018general}, that skew-scattering is proportional to the magnetix flux at high energies. However, the latter vanishes for a target skyrmion. So, we expect a greater suppression of the Hall effect at higher energies relative to that of, e.g., conventional skyrmion generating nonzero magnetic flux. This consideration indicates that the Hall effect in a target skyrmion is a nonmonotonic function of both the carrier energy and target skyrmion radius. In the rest of the paper, we delve into that analysis for more realistic target skyrmion geometries.  



\begin{figure}[hbt!]
    \centering  
     \includegraphics[width=1.0\linewidth]{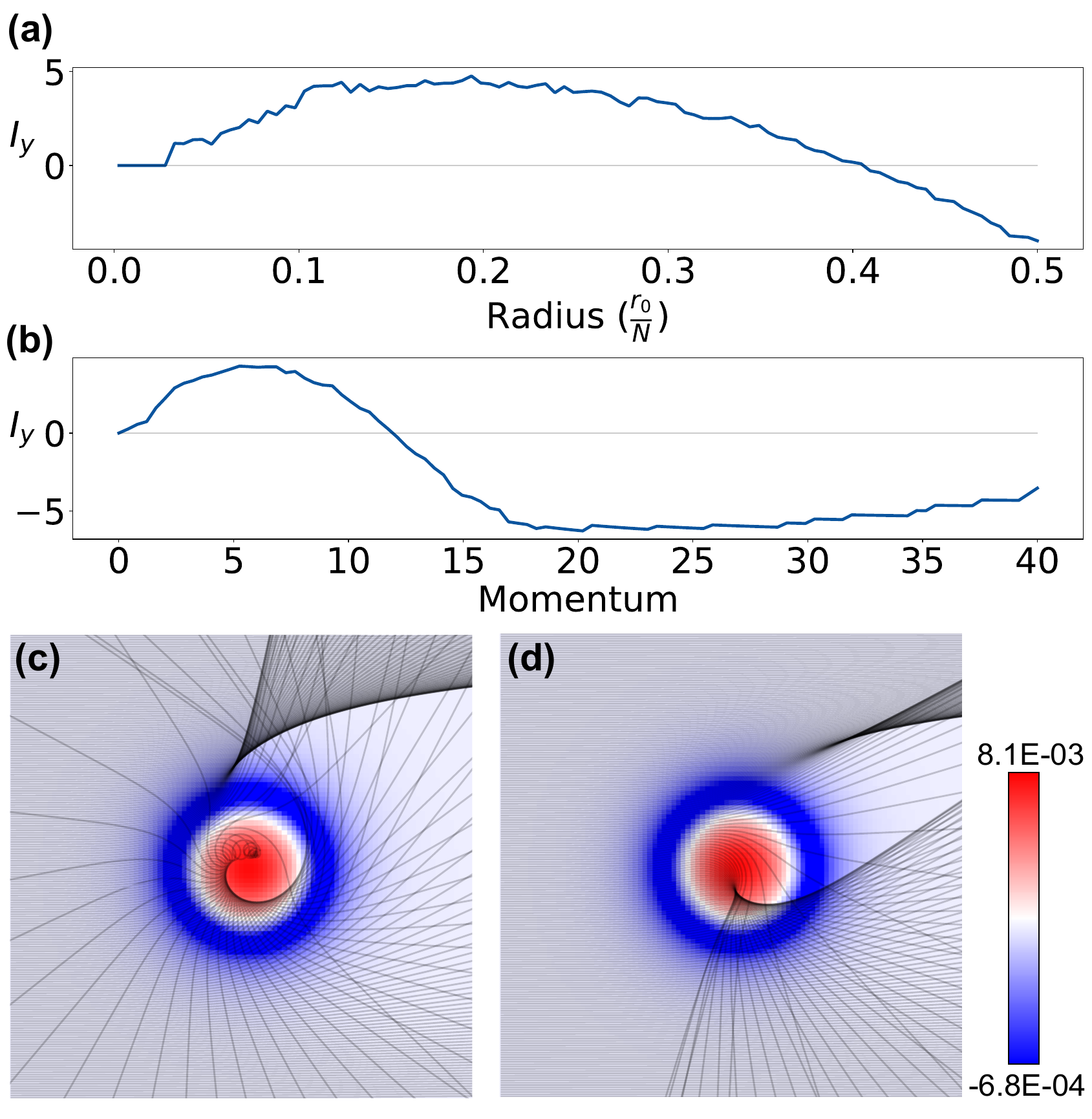}
     \caption{a) Transverse current as a function of target skyrmion radius for a fixed momentum $p = 15$. b) transverse current as a function of momentum for $r_0 = N/2$. c,d) Electron trajectories in a target skyrmion's effective field for $p = 5$ and $p = 15$.} 
     \label{fig:classical_trajectories}
\end{figure}

Next, we solve a continuous model of a target skyrmion profile illustrated in Fig.~\ref{fig:classical_trajectories}. We simulate a classical model of electrons traveling through a target skyrmion. Electrons obey Lorentz force $F = q(\textbf{E} + \textbf{v} \times \textbf{B})$, where there is no external electric field $\textbf{E} = 0$ in our model. Using $\textbf{B} = B_z(\textbf{r})$ from Fig.\ref{fig:target_skyrmion}b, we calculate the transverse current $I_\text{y} = \int d\rho$ $v_y(\rho)$. We used Runge-Kutta method to solve the differential equations of electrons in a magnetic field for $v_y(\rho)$. The initial condition for the simulation were an array of electrons uniformly distributed along the left lead with a fixed initial momentum in $\hat{x}$-direction. Here we set the electron charge and mass to one. 

We calculated $I_{\text{y}}$ as a function of target skyrmion radius (Fig. \ref{fig:classical_trajectories}a) and as a function of electron momentum (Fig. \ref{fig:classical_trajectories}b). A sign switch in $I_\text{y}(r_0)$ indicates a competing order between the inner and outer shell. The trajectories of the electrons traveling through the target skyrmion were plotted in Fig. \ref{fig:classical_trajectories}c,d for $p = 5$ and $p = 15$, respectively. For small momentum $p < 12$, $I_\text{y}$ is positive indicating that most of the electrons are deflected by the negative magnetic field of the outer shell. As shown in \ref{fig:classical_trajectories}c where $p = 5$, the trajectories ended up in the top lead more than than bottom lead. For $p > 12$, we observed a sign switch of $I_\text{y}$. Electrons with high momentum can pass through the outer shell with little change in their trajectories, but are strongly deflected to the bottom lead by the inner shell. In the high momentum limits, we see that $I_\text{y}$ approaches 0. Classically, we have shown that topological trivial target skyrmion can give raise to a non-monotonic behavior of the Hall effect due to variations in the magnetic field. The next section will focus on quantum transport of the target skrymion.

\section{Numerical modeling.} \label{sec:numerical}
\subsection{Tight binding model} \label{sec:tightbinding}
Here we used Kwant, a software package for quantum transport, to calculate the Hall resistance and Hall angle of target skyrmion \cite{groth2014kwant}. The tight-binding Hamiltonian of the conduction electron is given by,
\begin{equation}
    H = -t\sum_{\langle ij \rangle} c_{i}^{\dagger}c_j  - J\sum_{i} c_{i}^{\dagger} \bm{ \sigma} \cdot \bm{S} c_i,
\end{equation}
where $c_{i}^{\dagger}$ and $c_i$ are the fermionic creation and annihilation operators of an electron at site $i$, $t = 1$ is the nearest neighbor hopping, and $J$ is the Hund's coupling strength between the itinerant electrons and the background spin texture $\bm{S}$. The band is symmetric with a band width of $8t + 2J$ for $J < 4$. Above $J > 4$, the band is gapped with a gap of $2J - 8t$. Here we focus on $ J < 4t$, and the transport energy right below the edge of the band overlap region $E \approx 4t - J$. In this range, we observed a non-monotonic behavior of the Hall effect as a function of the target skyrmion radius (Fig. \ref{fig:transport}a), where it is monotonic for the skyrmion in Ref. \cite{ndiaye2017}.
\subsection{Hall resistance and Hall angle}\label{sec:Halleffect}
We simulate four ferromagnetic leads attached to the sample and calculate the transmission probabilities. The leads are fully attached to the left, right, top and bottom of the system. We impose current ($I$) from the left lead to the right lead and measure the voltage ($V$) across the top and bottom lead. We applied Landauer-B\text{\"u}ttiker formalism to obtain the Hall resistance and Hall angle. The equation for Hall resistance is computed as \cite{datta1997electronic}, 


\begin{equation}
    R_{\text{H}} = \frac{V}{I} = \frac{h}{2e^2} \frac{R^2 - L^2}{(R+L)(R^2 + L^2 + 2F(F+R+L))}, 
    \label{eq:RH}
\end{equation}
where we set $h/e^2$ to unity. See appendix.\ref{HallResistanceDerivation} for leads configuration and derivation of the Hall resistance.

The Hall angle is defined as the resultant angle between the Hall voltage ($E_H$) and the longitudinal voltage ($E_x$). Since $E_H$ contributed by the spin texture is very small compared to $E_x$, the Hall angle thus can be expressed as the ratio between $E_H$ and $E_x$ (Eq~\ref{eq:Hallangle}). Where $E_H$ ($E_x$) is the difference between the top and bottom terminal voltages (left and right terminal voltages). Detailed derivations of the terminal voltages can be found in Ref. ~\cite{ndiaye2017, yin2015topological}.

\begin{equation}
    \theta_{\text{H}} = \frac{E_H}{E_x} = \frac{V_T - V_B}{V_R - V_L}, 
    \label{eq:Hallangle}
\end{equation}

\section{Results} \label{sec:results}
\subsection{Transport results for a target skyrmion} \label{sec:targetSkyrmion}

\begin{figure}[hbt!]
    \centering
     \includegraphics[width=1.0\linewidth]{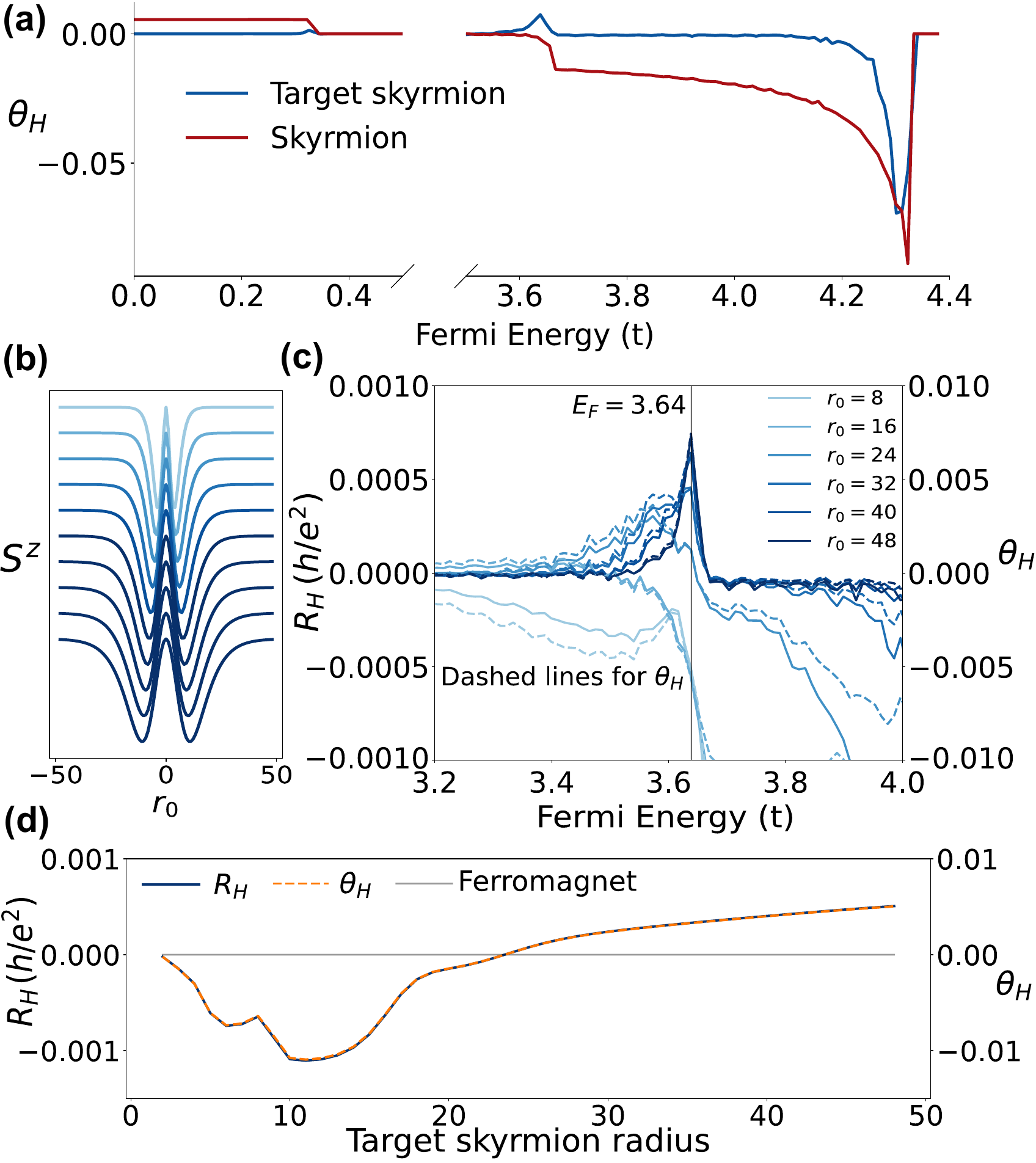}
     \caption{(a) Target skyrmion and skyrmion Hall angle as a function of energy for $N$x$N$ = $96\times96$, $r_0 = 48$, and $J = 1/3$. Broken plot to help focus on the non-zero Hall angle region. (b) $S^z$ profile taken across the center of the target skyrmion for various radii. The spins are rotated by $2\pi$ with respect to the center of spin. (c) Zoomed in plot for Hall resistance and Hall angle as a function of the transport energy for various target skyrmion radii.(d) The Hall resistance and Hall angle as a function of target skyrmion radius for $E = 3.64$.}
     \label{fig:transport}
\end{figure}

 Here we start with a big picture by comparing the Hall angles between the normal skyrmion and the target skyrmion for $J = 1/3$. We calculated the Hall angle for system size $N$x$N$ = 96$a_0$x96$a_0$ where $a_0 = 1$ is the lattice constant, and large initial $r_0 = 48$ to satisfy the adiabatic approximation for the skyrmion \cite{ndiaye2017}. In the next section we show that $r_0 = 48$ also satisfies the adiabatic approximation for the target skyrmion. By sweeping the energy from $E = 0$ to $E = 4t + J$, we observed a zero Hall angle for $|E| < J$ in the target skyrmion, contrasting to a constant Hall angle for the skyrmion. For $J < E < 4t - J$ where the spin up and spin down electrons are mixed due to the band overlap, the skyrmion has a zero Hall angle, while the target skyrmion has a peak at the edge of the band overlap where $E \approx 4t - J$. The zero Hall angle phenomena for the skyrmion has been thoroughly investigated in Ref.\cite{ndiaye2017,yin2015topological}. 
 
 
 We look closely at the peak at $E \approx 4t - J$ for the target skyrmion and study its radius dependence. Figure \ref{fig:transport}b shows $S^z$ along a line cut through the middle of the target skyrmion. As the radius increases, the target skyrmion transitions from a sharp spin texture to a slowly varying spins texture. Fig \ref{fig:transport}c shows the Hall angle and resistance for various radii. The peak disappears as the radius decreases. For small radius, $r_0 < 10$, the Hall angle at full energy range is similar to the skyrmion. We fixed the transport energy at the peak position $E = 3.64$ for $r_0 = 48$ and record the Hall angle and resistance.

Fig \ref{fig:transport}d shows a non-monotonic behavior of $R_\text{H}(r_0)$ and $\theta_{\text{H}}(r_0)$. For $r_0 = 0$, the system is in a perfect ferromagnetic state, therefore does not exhibit any Hall effects. For $r_0 < 24$, $R_\text{H}$ and $\theta_{\text{H}}$ are negative. We also see a rough behavior for $r_0 < 10$. This can be understood as when the inner shell is too small (i.e. its radius is less than 4$a_0$), it can hardly be considered as a skyrmion. Therefore, the system as a whole is one large skyrmion with spin defects at the center, which explains the negative Hall effect in this region. As the radius increases, the inner shell starts to expand into a skyrmion. The outer and inner shell scatter electrons in opposite directions, and eventually they both cancel out for $r_0 \approx 24$. For $r_0 > 24$, $R_\text{H}(r_0)$ and $\theta_{\text{H}}(r_0)$ are positive, indicating that Hall effect is contributed by the inner shell. For comparison with a skyrmion, the Hall resistance and Hall angle become constant and independent of skyrmion's radii for $r_0 > 8$ \cite{ndiaye2017}. Here, we show that the Hall effect sign switches from negative to positive owing to the topological Hall effect of the outer and inner shell, respectively.


\subsection{Adiabatic approximation: J dependence} \label{sec:multiTargetSkyrmion}


To demonstrate that the results respect the adiabatic approximation, we show that the adiabatic parameter $\lambda_a >> 1$, where $\lambda_a$ is defined as \cite{bruno2004topological, denisov2020theory}

\begin{equation}
    \lambda_a = \frac{\omega_s \xi}{v_F},
    \label{eq:adiabatic_parameter}
\end{equation}
where $v_F$ is the Fermi velocity electrons in the lead. $\omega_s = J/\hbar$ is the electron precession frequency. $\xi$ is the characteristic length such that $S^z$ is rotated by an angle of $\pi$. We calculated $\lambda_a$ for various exchange coupling at the edge of the band overlap region to confirm that $\lambda_a$ is greater than one. 


\begin{figure}[hbt!]
    \centering
     \includegraphics[width=1.0\linewidth]{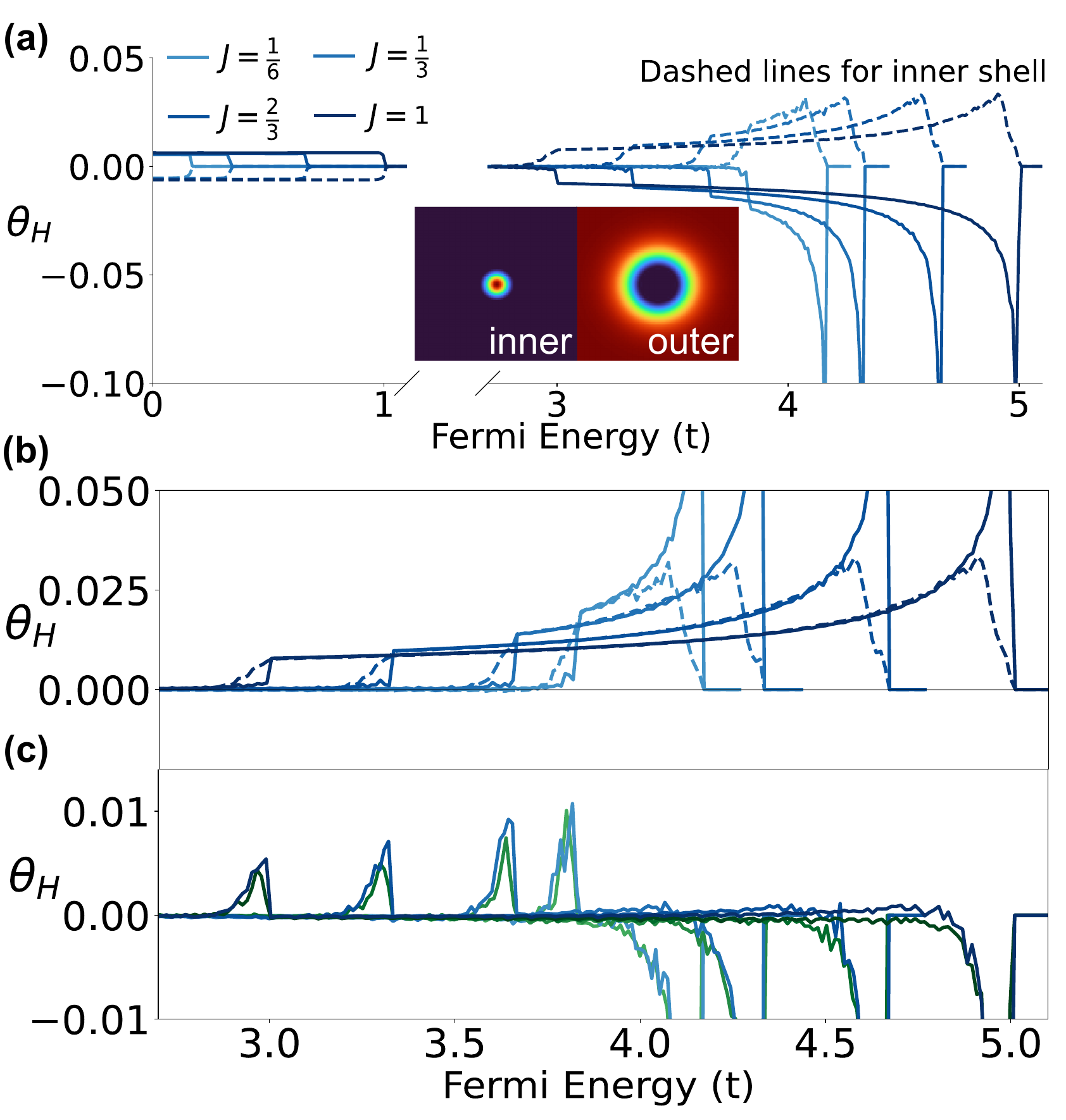}
     \caption{(a) Hall angle for the inner and outer shell, with parameters $r_0 = 48$ and $N = 96$. (b) Zoom in plot near the edge of the band overlap region, and plot $|\theta_{\text{H}}|$. (c) Hall angle plot of a target skyrmion (green plots) and combined Hall angle of the inner and outer shell (blue plots).}
     \label{fig:Jdependence}
\end{figure}


Without loss of generality, we focus only on the $\theta_{\text{H}}$ for the results presented in this section. We calculated the Hall angle of the target skyrmion for multiple $J$ with $J < 4t$, and observed the magnitude of Hall angles decrease as $J$ increases (Fig. \ref{fig:Jdependence}c). This may be seen as problematic as it indicates a vanishing Hall angle. Here we resolve this issue by demonstrating that the magnitude of the Hall angles are set by the finite Hall angle of a skyrmion in the adiabatic regime. To characterize $\theta_{\text{H}}$ dependence of the exchange coupling, we isolate the inner and outer shell of the target skyrmion and placed them in a ferromagnetic background with both having the same system size ($N = 96$). 

We placed the inner shell in a spin down ferromagnet to prevent scattering from an abrupt domain wall. Similarly, we placed the outer shell in a spin up ferromagnetic background. Fig. ~\ref{fig:Jdependence}a shows $\theta_{\text{H}}$ of the inner and outer shell, and the inset figures show $S^z$ of the inner and outer shell. Since the shells have opposite topological charge, their Hall angles also have the opposite signs: the inner shell has positive $\theta_{\text{H}}$ for $E > 4t - J$, and the outer shell has negative $\theta_{\text{H}}$. We zoom in on the energy range near the edge of the bands overlap and compare $|\theta_{\text{H}}|$ (Fig. \ref{fig:Jdependence}b). We observed both shells have the same magnitude for $4t - J < E$. $\theta_{\text{H}}$ of the inner shell has a soft decrease to zero starting from $E = 4t - J$, whereas the outer shell has an abrupt jump to zero. This difference gives raise to the Hall angle and the magnitude is set by the single skyrmion (Fig.\ref{fig:Jdependence}c). In the adiabatic limits, all skyrmion Hall angle becomes constant right above the band overlap region, and thus the target skyrmion's Hall angle never vanishes for $J < 4t$. We compared the total Hall angle obtained by calculating the contribution from each shells versus the calculation of one target skyrmion. Here we demonstrated that the Hall angle of the target skyrmion is the sum of the inner and outer shell's Hall angles, as the Hall angles superimposed one another in Fig.\ref{fig:Jdependence}c. Furthermore, our results respect the adiabatic approximation which mean that the Hall angle arises from real space topology and can be classified as the topological Hall effect despite it the spin texture has trivial topology.  
\section{Conclusion} \label{sec:conclusion}
 In this work, we investigated the Hall effect in a recently discovered target skyrmion spin configuration  \cite{tang2021magnetic,  kolesnikov2018skyrmionium, zhang2018real, shen2019current, seng2021direct}. The target skyrmion is characterized by a zero skyrmion (topological) charge.  We started from a highly-simplified semiclassical model of a target skyrmion and found that the Hall effect is a non-monotonic function of both the carrier energy and target skyrmion radius. On top of that, we found that the Hall effect changes sign, which indicates competition between opposite Hall effect produced by the inner and outer shell of the target skyrmion. This observation carries over to the fully quantum mechanical treatment, where we observed a nonmotonic Hall angle and Hall resistance as a function of starget skyrmion radius. We developed a method to better understand the transport contribution in the target skyrmion by isolating the shells in ferromagnetic backgrounds. We showed that, in the broad region of energies, the Hall effect can be decomposed into a superposition of the Hall effects produced by the distinct shells of the target skyrmion. This method could be applied to more complex spin textures, which may facilitate understanding of the result of transport theory calculations. Our results suggest that it may possible to detect the presence of target skyrmions in experiments and call for further experimental investigation of the Hall effect in such textures. Finally, our result challenges the popular belief in the community that the Hall effect necessarily implies a nonzero skyrmion number.  

\section{Acknowledgement}\label{sec:Acknowledgement}
This work was supported by the U.S. Department of Energy, Basic Energy Science under award No. DE-SC0020221.

\appendix

\section{Hall resistance}\label{HallResistanceDerivation}
\begin{figure}[hbt!]
    \centering
     \includegraphics[width=1\linewidth]{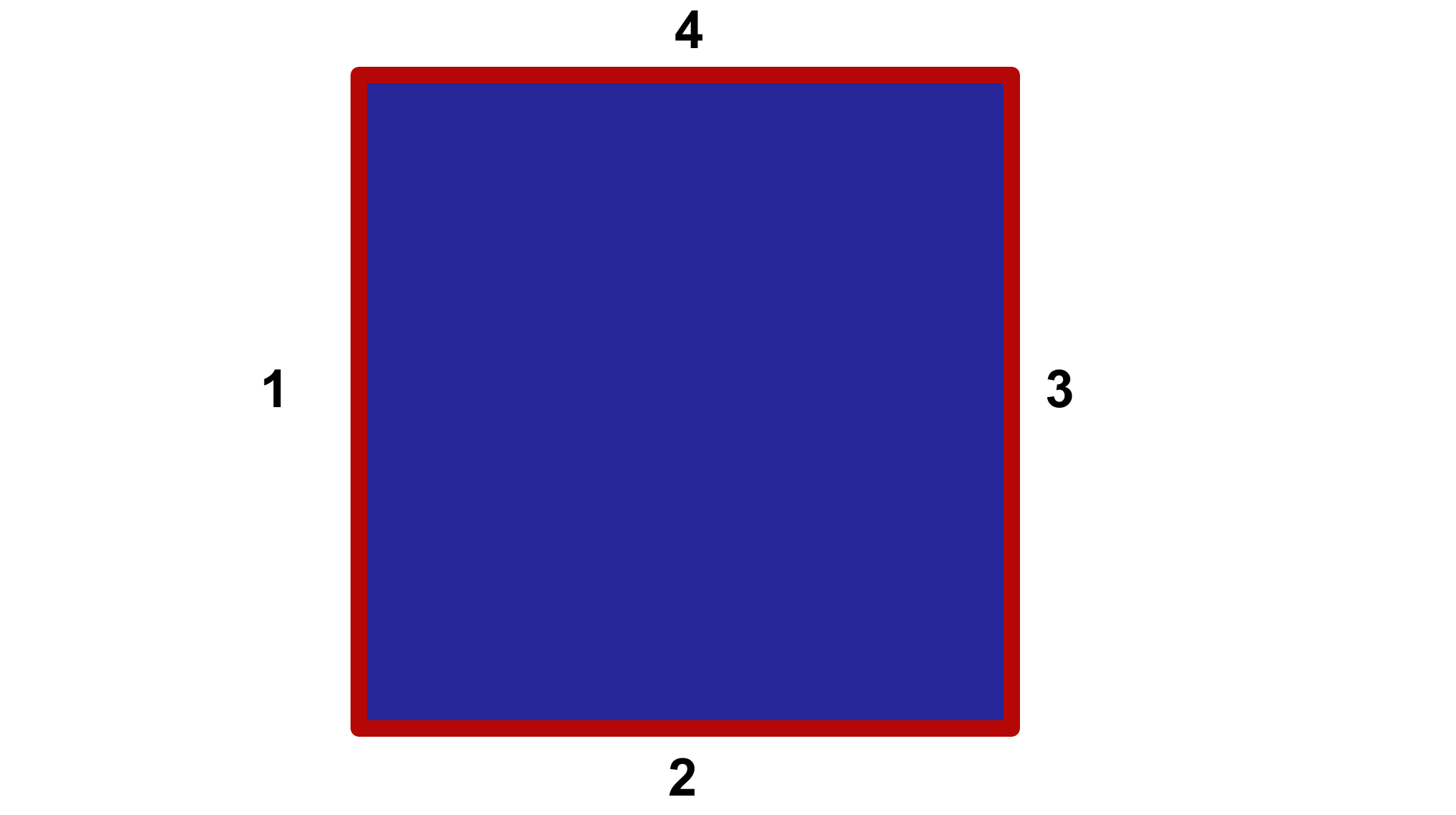}
     \caption{The leads (red) are fully attached to the scattering region (blue).}
     \label{fig:leads}
\end{figure}  
We ran a current from lead 1 to lead 3, and measure the voltage across lead 2 and 4. Without loss of generality we set $V_4 = 0$ and measure $V_2$. Fig. \ref{fig:leads} shows the leads configuration. We start with $\bm{I} = \bar{\bm{G}}\bm{V}$, where $\bar{\bm{G}}$ is the the conductance matrix defined as
\begin{equation}
    \bar{\bm{G}} = \frac{2e^2}{h}\begin{bmatrix}
                    T_{0} & -T_{12} & -T_{14}\\
                    -T_{21} & T_{0} & -T_{24} \\
                    -T_{41} & -T_{42} & T_{0}
                    \end{bmatrix} .
    \label{eq:condutance}
\end{equation}
Since four leads are completely symmetric, we can define forward transmission, right turning, and left turning as follows: 
\begin{equation}
\begin{matrix}
F = T_{13} = T_{31} = T_{24} = T_{42} \\
R = T_{12} = T_{23} = T_{34} = T_{41} \\ 
L = T_{14} = T_{43} = T_{32} = T_{21}
\end{matrix}
\label{eq:transmissionProb}
\end{equation}
 and 
 
\begin{equation}
T_0 = F + R + L
\label{eq:totalTransmission}
\end{equation}

Solve for the voltage $\bm{V} = \bar{\bm{R}}\bm{I}$ where $\bar{\bm{R}}$ is the resistance matrix (i.e. inverse of the conductance matrix). Using the current $I_1 = I = -I_3$ and $I_2 = 0$, we obtained

\begin{equation}
V_2 = \frac{I}{|\bar{\bm{G}}|}(R^2 - L^2) ,
\label{eq:V2}
\end{equation}
where
\begin{equation}
|\bar{\bm{G}}| = \frac{2e^2}{h}(R+L)(R^2 + L^2 + 2F(F + R + L))
\label{eq:detG}
\end{equation}
We arrive at the Hall resistance $R_\text{H} = V_2/I$ which is Eq.\ref{eq:RH}.

\newpage
\bibliography{biblio}			


\end{document}